\documentclass[prl,aps,twocolumn,groupedaddress,amsmath,showpacs]{revtex4}

\usepackage{graphicx}
\usepackage{dcolumn}
\usepackage{bm}

\begin{document}

\preprint{APS/123-QED}

\title{Superconducting PrOs$_{4}$Sb$_{12}$: a thermal conductivity study}

\author{ G. Seyfarth$^{1}$, J.P. Brison$^{1}$, M.-A. M{\'e}asson$^{2}$, D. Braithwaite$^{2}$, G. Lapertot$^{2}$, J. Flouquet$^{2}$}
\affiliation{$^{1}$CRTBT, CNRS, 25 avenue des Martyrs, BP166,
38042 Grenoble CEDEX 9, France}
\affiliation{$^{2}$DRFMC, SPSMS, CEA Grenoble, 38054 Grenoble,
France}

\date{\today}

\begin{abstract}
The superconducting state of the heavy fermion PrOs$_{4}$Sb$_{12}$
is studied by heat transport measurements on a highly homogeneous
single crystal exhibiting only one transition peak in the specific
heat. The field \textit{and} temperature dependence of the thermal
conductivity confirm multiband superconductivity and point to
fully open gaps on the whole Fermi surface.
\end{abstract}

\pacs{71.27.+a, 74.25.Fy, 74.25.Op, 74.45.+c, 74.70.Tx}

\maketitle

Several unusual features have been reported on the superconducting
state of PrOs$_{4}$Sb$_{12}$, the first Pr-based heavy fermion
(HF) superconductor \cite{Bauer I}: a double superconducting
transition in the specific heat has been observed on different
samples \cite{Maple,Vollmer,Marie-Aude,Grube}, like in the
well-known case of UPt$_{3}$, but its intrinsic nature has not
been clearly established yet \cite{Marie-Aude,Marie-Aude2006}.
Other experiments also point to \textit{unconventional}
superconductivity in this compound: thermal conductivity
($\kappa$) measurements in a rotated magnetic field \cite{Koichi},
London penetration depth studies \cite{Chia} and flux-line lattice
distortion \cite{Huxley2004} support nodes of the gap. These
results contrast with scanning tunneling spectroscopy (STM)
\cite{Hermann2004},  Sb nuclear quadrupole resonance (NQR)
\cite{Kotegawa}  or muon spin relaxation ($\mu$SR)
\cite{MacLaughlin} which measured a fully opened gap. Our first
very low temperature $\kappa$ measurements under magnetic field
(sample A) provided strong evidence for multiband
superconductivity (MBSC) in PrOs$_{4}$Sb$_{12}$, but sample
quality did not allow analysis of the gap topology from the low
temperature regime of $\kappa$ \cite{prl2005}.

In this Letter, we report a new study of thermal transport at very
low temperatures on a highly homogeneous PrOs$_{4}$Sb$_{12}$
single crystal. Supplementary specific heat measurements of this
sample show only one single, sharp superconducting jump at
$T_{c}$, supporting an extrinsic origin for the double transition
reported so far. Improved sample quality also has profound impact
on thermal transport, mainly on the temperature dependence
$\kappa(T)$ in zero field. It provides compelling evidence for a
rather "conventional" MBSC scenario with fully opened gaps on the
whole Fermi surface.

\begin{figure}[t]
\includegraphics[width= 8cm]{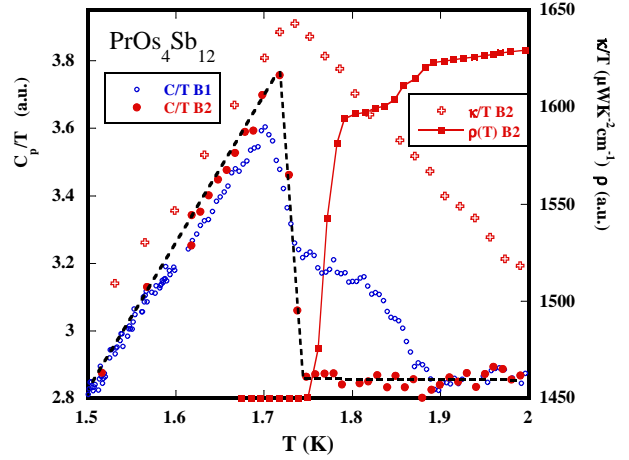}\caption{Specific heat $C_{p}/T$, electric resistivity
$\protect\rho$ and thermal conductivity $\protect\kappa(T)/T$ in
zero field around $T_{c}$, documenting the collapse of the double
transition (observed only in sample B1) and the high homogeneity
of sample B2. Nevertheless, a kink at 1.9~K on $\rho(T)$ reveals
some "traces" of the upper transition in sample B2, but the bulk
superconducting transition clearly corresponds to the lower $T_{c}$ in $C_{p}/T$(T) of sample B1.}%
\label{compaech}%
\end{figure}

Our thin, platelet-shaped PrOs$_{4}$Sb$_{12}$ single crystal
(sample B2, $\sim 760\times340\times45$~$\mu$m$^{3}$,
$T_{c}\simeq1.75$~K, residual resistivity ($\rho$) ratio
$\rho(300~K)/\rho(T_c)\sim30$ instead of $\sim15$ in sample A) has
been extracted (gently "grinded down" against the disk of a
diamond saw) from a conglomerate of several small cubes of
PrOs$_{4}$Sb$_{12}$ (sample B1
$\sim1\times0.75\times0.6$~mm$^{3}$), grown by the Sb-flux method
\cite{Koichi}. Specific heat ($C_p$) in zero field has been
measured on a PPMS, first the entire conglomerate (B1) and then
sample B2 alone (see fig. \ref{compaech}). $\protect\kappa(T,H)$
of sample B2 parallel to the magnetic field has been measured in a
dilution refrigerator by a standard two-thermometers-one heater
steady-state method down to 30~mK and up to 3~T
($\mu_{0}H_{c2}(T\rightarrow0)\simeq2.2$~T). The carbon
thermometers were thermalized on the sample by gold wires, held by
silver paint on gold stripes evaporated on the surface of the
sample after ion gun etching. The gold stripes are essential for
the stability and the quality of the electrical contacts
(resistance $R_{c}^{e}\approx 10\operatorname{m\Omega}$ at 4~K).
The same contacts and gold wires were used to measure the electric
resistivity of the sample by a standard four-point lock-in
technique. The reliability of the experimental setup was checked
against the Wiedemann-Franz law, giving quantitatively similar
results for $L/L_{0}$ as reported in \cite{prl2005}.

The excellent homogeneity of our sample B2 is documented by the
fact that the bulk superconducting transition appears at exactly
the same temperature on $C_{p}$, $\kappa/T$ and\ $\rho$ (see fig.
\ref{compaech} and below). Another criterion regarding crystal
purity is the residual value of $\kappa/T$ in the $T\rightarrow0$
limit. For platelet B2, it is smaller than $1.6\operatorname{\mu
W}\operatorname{K}^{-2}\operatorname{cm}^{-1}$, which corresponds
to $0.07\%$ of $\kappa/T(T\rightarrow0,
\mu_{0}H=2.5\operatorname{T}>H_{c2})$ and is significantly lower
than in former sample A. These signatures of high sample quality
allow us to use thermal transport at very low temperatures as a
sensitive probe of the low lying energy excitations in
PrOs$_{4}$Sb$_{12}$.

Figure \ref{compaech} displays the superconducting transition at
zero magnetic field as seen by specific heat for samples B1 and
B2, and in addition for sample B2 the corresponding $\kappa/T$ and
$\rho(T)$ curves. The specific heat of sample B1 exhibits a double
transition, comparable to those reported in \cite{Marie-Aude}. The
fact that both transitions behave similarly under magnetic field,
and that the upper one always appears inhomogeneous
\cite{Marie-Aude} had cast doubt on the intrinsic nature of this
double transition. The remarkable result is that the double
transition collapses to a single, sharp jump of $C_{p}$ (at the
lowest $T_{c}$ and of about the same overall height) just by
reducing the crystal dimensions (sample B1~$\rightarrow$~B2).
Obviously, areas with a single and a double transition coexist
within the same sample. It suggests that, like in
URu$_{2}$Si$_{2}$ \cite{Ramirez}, the observed double superconducting transition in PrOs$_{4}%
$Sb$_{12}$ is related to sample inhomogeneity. A hint for
extracting samples with a single transition comes from the
preparation stage: in order to remove all small cavities appearing
during the sawing process of B1, we had to reduce the thickness of
B2 down to only $50$~$\mu$m. Similarly, such tiny dimensions were
reported for another sample exhibiting a single $C_{p}$ jump
\cite{Marie-Aude2006}. Further systematic (structural)
investigations are required to determine the nature of defects
which might be at the origin of the sharp double transition in
PrOs$_{4}$Sb$_{12}$.

\begin{figure}[t]
\includegraphics[width= 8cm]{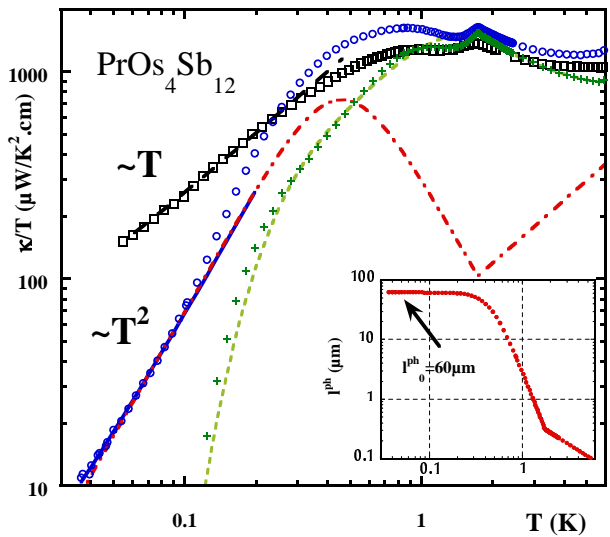}\caption{$\protect\kappa/T(T)$ in zero field: comparison of former
(A, open squares) and present (B2, open circles) sample. For
sample B2, we separate phonon $\kappa^{ph}$ (dash-dotted line) and
electronic $\kappa^{el}$ (crosses) contribution. The dashed line
is a fit of $\kappa^{el}$ within the MBSC scenario, see text. The
inset shows the estimated temperature dependence of the phonon
mean free path.}
\label{kappa_charge}%
\end{figure}

In figure \ref{kappa_charge}, we compare the temperature
dependence of $\kappa/T$ in zero field of samples A (former) and
B2 (present). For B2, $\kappa/T$ has dropped by about two orders
of magnitude from $T_c$ down to 30~mK, with a clean $T^3$ behavior
of $\kappa$ below 100~mK. In sample A, the low temperature
behavior of  $\kappa/T$ is probably dominated by impurities,
defects and/or other inhomogeneities, resulting in a sort of
cross-over regime $\kappa/T\sim T$ \cite{prl2005}. By contrast, at
higher temperatures, $\kappa/T$ is qualitatively the same for both
samples, except that there is now a sharp peak in $\kappa/T$
precisely at the superconducting transition (as seen by $C_{p}$
and $\rho$). In sample A, these features were much broader and the
local maximum of $\kappa/T$ appeared below the resistive $T_{c}$
or the onset of the specific heat jump.

\begin{figure}[t]
\includegraphics[width=8cm]{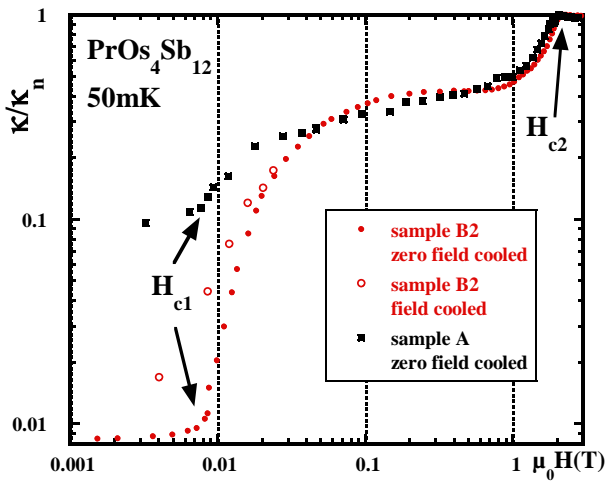}\caption{$\kappa/T(H)$ (normalized to its value at the superconducting transition) at 50~mK for samples A and
B. The arrows indicate the lower ($H_{c1}$) and upper ($H_{c2}$)
critical field. The data in the \textquotedblright field
cooled\textquotedblright\ mode reveal residual flux pinning below
$50~mT$ and a sensitivity of $\kappa$ to
fields as low as 5~mT.}%
\label{kappaH}%
\end{figure}

 In figure \ref{kappaH}, we compare the normalized $\kappa(H)/T$ data at
$50\operatorname{mK}$ of samples A and B2: in this temperature
region, the quasiparticle mean free path is governed by elastic
impurity scattering \cite{prl2005}. The very fast increase of
thermal conductivity at low fields is perfectly reproducible, and
even more pronounced in sample B2 because of the significant drop
of $\kappa/T$ for $T\ll T_{c}$ in zero field (see fig.
\ref{kappa_charge}): a magnetic field of only $5\%$ of $H_{c2}(0)$
is enough to restore about $40\%$ of the normal state heat
transport. This robust feature, similar to observations in
MgB$_{2}$ \cite{Sologubenko}, confirms MBSC in PrOs$_{4}$Sb$_{12}$
\cite{prl2005}.  The plateau at $0.4\kappa_n$ can be interpreted
as the "normal state" contribution of the small gap band at
$T\rightarrow 0$, observed once the vortex cores of that band
completely overlap. Moreover, despite improved sample quality,
testified by the sharp change of slope at $T_c$, there is clearly
no sign of a phase transition in the mixed state as suggested in
\cite{Koichi}: $\kappa(H,T\rightarrow0)$ has no anomaly at the proposed
$H^{\ast}$ line ($H^{\ast}(T\rightarrow0)\approx0.8~T$), whereas
the $B\rightarrow C$ transition in UPt$_3$ was clearly seen on
$\kappa(H)$ \cite{Hermann}).

Because of improved sample homogeneity, it is now worth  analyzing
the temperature behavior of $\kappa/T$. Just above $T_{c}$,
$L/L_{0}\lesssim1$ \cite{prl2005}, which indicates a phonon
thermal conductivity ($\kappa^{ph}$) negligible compared to the
electronic heat transport ($\kappa^{el}$) in the neighborhood of
$T_c$. The change of slope observed at $T_c$
($d(\kappa_{S}/\kappa_{N})/d(T/T_{c}))$ is of order 1.4. In
conventional superconductors, it is generally ascribed to the
combined effects of the opening of the gap and the energy
dependence of the electron-phonon scattering rate on
$\kappa^{el}$\cite{Beyer_Nielsen,Ambegaokar}. In the BCS
weak-coupling limit,  its maximum value is of order $1.4$, when
lattice scattering is the limiting mechanism for $\kappa^{el}$
(see measurements on very pure In \cite{Jones,Tewordt,Kadanoff}).
For PrOs$_4$Sb$_{12}$, electronic inelastic scattering may replace
the effect of electron-lattice scattering. Nevertheless, taking
into account the relative weight of elastic to inelastic
scattering, as well as MBSC (negligible effect of gap opening in
the small gap band), the value of
$d(\kappa_{S}/\kappa_{N})/d(T/T_{c})\approx 1.4$ appears very
large. It is likely a signature of strong-coupling effects, as
observed (and calculated) for example in lead
($d(\kappa_{S}/\kappa_{N})/d(T/T_{c})\approx 7$
\cite{Jericho,Beyer_Nielsen}). Indeed, Sb NQR \cite{Kotegawa} or
heat capacity analysis \cite{Grube} have already stressed
strong-coupling effects in PrOs$_4$Sb$_{12}$.

As regards now the origin of the local maximum of $\kappa/T$
slightly below 1~K, it had been ascribed to enhanced inelastic
mean free path due to the condensation of electronic scattering
centers for $T<T_{c}$. But the question of phonon or electronic
origin (or both) was left open \cite{prl2005}. It is now possible
to give a quantitative estimate of the phonon contribution:
indeed, a maximum of  $\kappa/T$ below $T_c$ followed by a $T^3$
behavior of $\kappa$ at low temperature is well-documented from
superconducting Pb and Nb \cite{Jericho,Connolly}, from the rare
earth nickel borocarbides RNi$_{2}$B$_{2}$C (R=Lu, Y)
\cite{BoakninPhysicaC,Sera,Hennings} and many other materials.

On cooling the phonon mean free path $l^{ph}$ increases from a law
$l^{ph}\sim T^{-1}$ (when it is limited by electron-phonon
interactions above $T_c$) up to a typical crystal dimension at the
lowest temperature. But an intermediate regime, described
empirically by $l^{ph}\sim T^{-1}(T_c/T)^n$, starts below $T_c$
due to the reduction of scattering by electrons (see inset of fig.
\ref{kappa_charge}). Quantitatively, an estimation for
$\kappa^{ph}/T$ is plotted in fig. \ref{kappa_charge} from:
\begin{equation}
\frac{1}{\kappa_{sc}^{ph}/T}=\frac{1}{\kappa_{normal}^{ph}/T}\left(
\frac {T}{T_{c}}\right)  ^{n=3}+\frac{1}{bl_{0}^{ph}T^{2}},
\label{eqlpm}
\end{equation}
We used $\kappa_{normal}^{ph}=aT^{2}$ with a=60~$\mu$W/K$^{3}$cm
(from deviation of the Lorenz number above $T_c$ \cite{prl2005}),
$bl_{0}^{ph}$ is fixed by $\kappa/T$ at lowest temperatures
($T<100\operatorname{mK}$), yielding $l_{0}^{ph}\sim60$ $\mu$m
($b=10.9\times10^{3}\operatorname{W}\operatorname{K}^{-4}\operatorname{m}^{-2}$).
$l_{0}^{ph}$ is of order the smallest crystal dimension. The
adjustable parameter is mainly the power law (n) for the boosted
temperature dependence of $l^{ph}$ : it proved impossible
(adjusting n) to account for the local maximum only by the phonon
contribution, so that a situation similar to that of Pb
\cite{Jericho} or the high-T$_c$ oxydes \cite{Cohn,Krishana,Yu} is
recovered. In any case, it is seen that $\kappa^{ph}$ 
should follow a $T^{3}$ behavior up to 0.3~K, giving, in the
temperature range 0.1-0.3~K, a robust estimate of the electronic
contribution $\kappa^{el}/T$.

Nevertheless, we tried to understand the normalized electronic
contribution $\kappa^{el}/\kappa_{2.5\operatorname{T}}^{el}$ up to
$T\leq0.6\operatorname{K}$ (region with dominant elastic impurity
scattering). The most striking feature is that it is not possible
to fit $\kappa^{el}(T)$ if one assumes a BCS gap corresponding to
$T_{c}=1.729\operatorname{K}$: $\kappa ^{el}/T$ starts to rise at
much lower temperatures than expected, requiring a smaller gap
value. Of course, this cannot be compensated by strong coupling
effects (which only make it worse, increasing the ratio
$\Delta/T_c$), nor by another estimation of the phonon
contribution ($\kappa^{ph}(T)$ cannot be larger than
$bl_{0}^{ph}T^{3}$, which is constrained by the measurements below
0.1~K). The data points can be quantitatively reproduced within a
MBSC scenario, when we include a small $\Delta_{\text{s}}(T)$ and
a large $\Delta_{\text{l}}(T)$ gap function with the same $T_{c}$,
and two associated conduction channels:
$\kappa^{el}/T=n_{s}\cdot\kappa_{\Delta_{s}}^{el}/T+(1-n_{s})\cdot
\kappa_{\Delta_{l}}^{el}/T$. The best data fit is then obtained
for a zero temperature gap ratio of about
$\Delta_{\text{l}}/\Delta_{\text{s}}(T\rightarrow0)\sim3$ with
$\Delta _{\text{s}}(T\rightarrow0)\sim1\operatorname{K}$, and a
 \textquotedblright weight\textquotedblright\ for the small gap band $n_{s}\sim0.35$. This value is close to the
40\% deduced from the "plateau" of $\kappa(H,T\rightarrow 0)$
(fig. \ref{kappaH}). The characteristic field
scale $H_{c2}^{S}$ for the vortex core overlap of the small band gap can now be estimated from $H_{c2}/H_{c2}^{S}%
\sim\left(  \frac{\Delta_{\text{l}}\cdot\text{v}_{\text{s}}^{F}}%
{\Delta_{\text{s}}\cdot\text{v}_{\text{l}}^{F}}\right)  ^{2}$,
where $v^F_i$ is the average Fermi velocity of band i. In
\cite{Marie-Aude} we assumed that the small gap band is also a
light carrier band, with v$_{\text{F,s}}/$v$_{\text{F,l}}\sim5$.
We then get $H_{c2}^{S}\sim10\operatorname{mT}$, which is of the
order of $H_{c1}$ and seems reasonable owing to the data of
$\kappa(H)$. So the main outcome of the analysis of this new data
is the existence of a small but finite gap $\Delta_{\text{s}}$ in
PrOs$_{4}$Sb$_{12}$, quantitatively consistent with the MBSC
scenario deduced from $\kappa(H)$.

\begin{figure}[t]
\includegraphics[width=8cm]{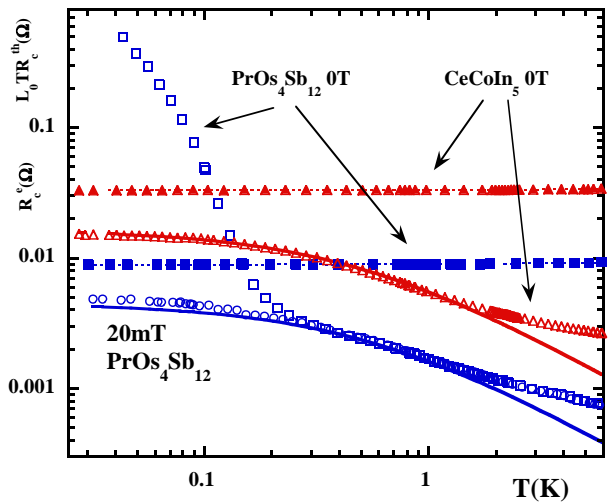}
\caption{Electric $R_{c}^{e}(T)$ (full symbols) and thermal $R_{c}^{th}(T).L_0T$ (open symbols) resistance of the Au-PrOs$_{4}$Sb$_{12}$ or Au-CeCoIn$_{5}$
contacts. Full lines: fits of $R_{c}^{th}(T)$ according to equation (\protect{\ref{eqRcth}}). $R_{c}^{e}(T)$ are temperature and field independent, but $R_{c}^{th}(T)$ shows a highly singular behavior in zero field below 200mK on PrOs$_{4}$Sb$_{12}$. This further supports  a fully open gap in this system (see text).}
%
\label{Rcontact}%
\end{figure}

Various measurements on the superconducting state of
PrOs$_{4}$Sb$_{12}$ have been interpreted either as pointing to
gap nodes \cite{Koichi,Chia,Huxley2004} or fully open gaps
\cite{Kotegawa,MacLaughlin,Hermann2004}. Focusing on the latest,
we can compare the extracted gap values. The NQR \cite{Kotegawa}
as well as the $\mu$SR \cite{MacLaughlin} studies propose large
ratios of $2\Delta/k_{B}T_c$, respectively $\sim5.2$ and
$\sim4.2$, supporting strong coupling effects but not the presence
of a small gap. Nevertheless, the NQR data shows a large residual
relaxation rate ($1/T_1$) below 0.5~K, which may point, as for our
sample A, to crystal inhomogeneities which prevent observation of
the smallest gap. Moreover, like specific heat, the nuclear
relaxation rate should be rather sensitive to bands with large
density of states. So if our interpretation \cite{prl2005} of the
small gap band as being also a "light mass" band is correct, it
may have indeed little contribution to $1/T_1$. The muon
relaxation rate ($\sigma_s$) measured with $\mu$SR is controlled
by the field distribution which may not put more weight on the
heavy than on the light bands. But the measurements were performed
in a residual field of 20~mT, already larger than $H_{c2}^{S}$, so
that again $\sigma_{s}(T)$ is probably governed by the high energy
excitations. But the "unusual" non linear field dependence of
$\sigma_{s}$($T=0.1$~K) compares well with $\kappa(H)$: the MBSC
scenario, with the small gap band associated to light carriers can
even "explain" the increase of $\sigma_s$ at low fields as
$\sigma_{s}\propto 1/m^{\ast}$. Eventually, the STM measurements
\cite{Hermann2004} proposed a gap distribution, which may extend
from 120~$\operatorname{\mu V}$ to 325~$\operatorname{\mu V}$
($2\Delta/k_{B}T_{c}\sim$1.5-4.1), not so far from our analysis of
$\kappa(T)$ ($2\Delta_{S}/k_{B}T_{c}\sim1.15$,
$2\Delta_{L}/k_{B}T_{c}\sim3.5$).

A further robust and reproducible \cite{prl2005} experimental observation supports a  \textit{fully open} gap in PrOs$_{4}$Sb$_{12}$. Indeed, we can also measure on our setup the electrical contact resistance ($R_{c}^{e}(T)$) between the sample and the gold wire of the thermometer thermalisation, as well as its
\textit{thermal} contact resistance $R_{c}^{th}$.  $R_{c}^{th}$ is defined as $P/\Delta T$, with P the Joule power dissipated in $R_{c}^{e}(T)$ and $\Delta T$ the thermal gradient produced by $P$. For large contact areas, we expect that at low temperature
\begin{equation}
R_{c}^{th}\approx 1/2\frac{1}{\frac{L_0T}{R_{c}^{e}}+aT^2}
\label{eqRcth}
\end{equation}
($aT^2$ being the phonon contribution to the thermal conductivity
of the contact). Figure \ref{Rcontact} shows that this is well
observed below 1~K for  PrOs$_{4}$Sb$_{12}$ in a field of 20~mT,
or CeCoIn$_5$ in zero field (new measurements on this same setup).
But for PrOs$_{4}$Sb$_{12}$ in zero field, an unexpected
divergence is observed below 200~mK, with no correspondence on
$R_{c}^{e}(T)$ (field and temperature independent). On the same
footing, CeCoIn$_{5}$ in zero field, like PrOs$_{4}$Sb$_{12}$ in
20~mT, could be cooled below 10~mK whereas PrOs$_{4}$Sb$_{12}$ in
zero field remained stuck above 25~mK. A reason could be that
electronic heat transport is suppressed at the
normal-superconducting interface when $k_{B}T\ll\Delta_{\text{s}}$
(whereas electric current can always go through thanks to Andreev
processes \cite{Andreev}): of course, this barrier is suppressed
in very low field in PrOs$_{4}$Sb$_{12}$ (multiband effects) or in
zero field in CeCoIn$_5$ which has line nodes of the gap
\cite{mov}.


In conclusion, we measured the thermal conductivity of a highly
homogeneous PrOs$_{4}$Sb$_{12}$ single crystal exhibiting a
\textit{single} jump on $C_{p}$ at $T_{c}$. The reproducible field
dependence $\kappa(H)$ at $T\ll T_{c}$ confirms the proposed MBSC
scenario. Further support now comes from the low temperature
$\kappa(T)$ and $R_{c}^{th}(T)$ data which both point to
isotropic, fully opened gap functions, in comparison with other
measurements (NQR, $\mu$SR, STM). Owing to the still mysterious
homogeneity problems in this compound and to the strong field
sensitivity, analysis of the data requires a close look at the
experimental conditions (sample, field and temperature range)
which may explain the remaining divergences of interpretations.

\begin{acknowledgments}
We are grateful for stimulating discussions with M. Zhithomirsky,
K. Izawa and H. Suderow, and for practical help and advice to A.
De Muer. Work supported in part by grant ANR-ICENET NT05-1\_44475.
\end{acknowledgments}


\begin{thebibliography}{99}


\bibitem {Bauer I}E.D. Bauer \textit{et al.}, Phys. Rev. B \textbf{65},
100506(R) (2002)

\bibitem {Maple}M.B. Maple\textit{ et al.}, J. Phys. Soc. Jpn. \textbf{71}
Suppl., 23 (2002)

\bibitem {Vollmer}R. Vollmer\textit{ et al.}, Phys. Rev. Lett. \textbf{90},
057001 (2003)

\bibitem {Marie-Aude}M.A. M\'{e}asson\textit{ et al.}, Phys. Rev.\textit{ }B
\textbf{70}, 064516 (2004)

\bibitem {Grube}K. Grube \textit{et al}., Phys. Rev.\textit{ }B \textbf{73},
104503 (2006)

\bibitem {Marie-Aude2006}M.A. M\'{e}asson\textit{ et al.}, Physica B, \textbf{378-380}, 56 (2006)

\bibitem {Koichi}K. Izawa \textit{et al.}, Phys. Rev. Lett. \textbf{90},
117001 (2003)

\bibitem {Chia}E. E.M. Chia \textit{et al.}, Phys. Rev. Lett. \textbf{91},
247003 (2003)

\bibitem {Huxley2004}A. Huxley \textit{et al.}, Phys. Rev. Lett. \textbf{93},
187005 (2004)

\bibitem {Hermann2004}H. Suderow \textit{et al.}, Phys. Rev. B \textbf{69},
060504(R)(2004)

\bibitem{Kotegawa} H. Kotegawa \textit{et al.}, Phys. Rev. Lett.
\textbf{90}, 027001 (2003)

\bibitem{MacLaughlin} D.E. MacLaughlin \textit{et al.}, Phys. Rev. Lett.
\textbf{89}, 157001 (2002)

\bibitem {prl2005}G. Seyfarth\textit{ et al.}, Phys. Rev. Lett. \textbf{95},
107004 (2005)

\bibitem {Ramirez}A.P. Ramirez \textit{et al.}, Phys. Rev.\textit{ }B
\textbf{44}, 5392 (1991)

\bibitem {Sologubenko}A.V. Sologubenko\textit{ et al.}, Phys. Rev. B
\textbf{66}, 014504 (2002)

\bibitem {Hermann}H. Suderow \textit{et al.}, J. Low Temp. Phys. \textbf{108},
11 (1997)

\bibitem {Beyer_Nielsen}J. Beyer Nielsen \textit{et al.}, Phys. Rev. Lett.
\textbf{49}, 689 (1982)

\bibitem {Ambegaokar}V. Ambegaokar \textit{et al.}, Phys. Rev. \textbf{139},
A1818 (1965)

\bibitem {Jones}R.E. Jones \textit{et al.}, Phys. Rev. \textbf{120}, 1167 (1960)

\bibitem {Tewordt}L. Tewordt, Phys. Rev. \textbf{128}, 12 (1962)

\bibitem {Kadanoff}L.P. Kadanoff \textit{et al.}, Phys. Rev. \textbf{124}, 670 (1961)

\bibitem {Jericho}M.H. Jericho\textit{ et al.}, Phys. Rev. B
\textbf{31}, 3124 (1985)

\bibitem {Connolly}A. Connolly \textit{et al.}, Proc. Roy. Soc. (London)
\textbf{A 266}, 429 (1962)

\bibitem {Sera}M. Sera\textit{ et al.}, Phys. Rev. B \textbf{54}, 3062 (1996)

\bibitem {BoakninPhysicaC}E. Boaknin\textit{ et al.}, Physica \textit{ }C
\textbf{341-348}, 1845 (2000)

\bibitem {Hennings}B.D. Hennings\textit{ et al.}, Phys. Rev. B \textbf{66},
214512 (2002)

\bibitem {Cohn}J.L. Cohn\textit{ et al.}, Phys. Rev. B \textbf{45},
13144 (R) (1992)

\bibitem {Krishana}K. Krishana \textit{et al.}, Phys. Rev. Lett. \textbf{75},
3529 (1995)

\bibitem {Yu}R.C. Yu\textit{ et al.}, Phys. Rev. Lett. \textbf{69}, 1431 (1992)


\bibitem {Andreev}A.F. Andreev, J. Exptl. Theoret. Phys. (U.S.S.R.) \textbf{46}, 1823 (1964), Sov. Phys. JETP \textbf{19}, 1228 (1964).





\bibitem {mov}R.~Movshovich \textit{et al.}, Phys. Rev. Lett. \textbf{86},
5152 (2001)
\end{thebibliography}
\end{document}